\documentclass[twocolumn,aps,pra,a4paper,10pt,showpacs,nofootinbib,preprintnumbers]{revtex4-1}

\usepackage{graphicx}
\usepackage{dcolumn}
\usepackage{bm}%
\usepackage{amsmath,amsfonts,amssymb}
\usepackage{graphicx}

\providecommand{\bra}[1]{\left\langle #1 \right|}
\providecommand{\ket}[1]{\left| #1 \right\rangle}

\providecommand{\fabs}[1]{\left| #1 \right|}

\providecommand{\mean}[1]{\langle #1 \rangle}
\providecommand{\grad}[1]{{\bm\nabla}}

\DeclareMathOperator{\tr}{tr}

\DeclareMathOperator{\spec}{spec}

\def\vechat#1{\text{\bfseries{\^#1}}}
\def\vec#1{\text{{\bfseries#1}}}

\begin{document}

\title{Entanglement-fluctuation relation for bipartite pure states}

\author{Aura Mae B. Villaruel}
 \email{avillaruel@nip.upd.edu.ph}
\author{Francis N.~C.~Paraan}
 \email{Corresponding author: fparaan@nip.upd.edu.ph}
\affiliation{National Institute of Physics, University of the Philippines Diliman, 1101 Quezon City, Philippines}


\date{\today}

\begin{abstract}
\noindent We identify subsystem fluctuations (variances) that measure entanglement in an arbitrary bipartite pure state. These fluctuations are of observables that generalize the notion of polarization to an arbitrary $N$-level subsystem. We express this polarization fluctuation in terms of subsystem purity and other entanglement measures. The derived fluctuation-entanglement relation is evaluated for the ground states of a one-dimensional free fermion gas and the Affleck-Kennedy-Lieb-Tasaki spin chain. Our results provide a framework for experimentally measuring entanglement using Stern--Gerlach-type state selectors.
\end{abstract}



\maketitle

\section{Introduction}
That there is a strong connection between bipartite entanglement and subsystem fluctuations is well-known. Previous studies on bipartite ground states of critical spin chains revealed that subsystem fluctuations in a conserved charge have the same (logarithmic) scaling with subsystem size as the entanglement entropy \cite{song2010a,laflorencie2015a}. Further investigations on one-dimensional free fermion systems have shown that R\'enyi entanglement entropies can be expressed in terms of the cumulants of the fermion number distribution within a partition \cite{klich2009a,song2012a,calabrese2012a}. Additionally, the link between entanglement and fluctuations has been studied in the context of cosmological models \cite{nambu2008a,kanno2014a} and topological invariants \cite{zhang2014a}. 

Simple qualitative arguments can provide insight into the deep relationship between fluctuations and entanglement. When a pure state is partitioned into two entangled subsystems, the state of either cannot be specified by a single state vector. The complete description of a subsystem (say $q$) requires  a  reduced density operator $\rho_q$ that is mixed \cite{feynman1963a}. The statistical uncertainty in determining the state of this partition manifests as a non-zero entanglement entropy, as well as non-zero fluctuations in subsystem observables that originate from inter-partition quantum correlations. In fact, the range of these correlations has been demonstrated to be a crucial factor in determining whether an entanglement area law will hold or not \cite{wolf2008a,eisert2010a,brandao2013a,frerot2015a}.

Entanglement in a pure state $\rho$ that is partitioned into a subsystem $q$ and bath $B$ is completely characterized by the entanglement spectrum \cite{li2008a} that consists of the eigenvalues of the reduced density operator $\rho_q = \tr_B \rho$.  Due to normalization and the Hermiticity of $\rho_q$, this spectrum has $N-1$ independent real parameters when $\rho_q$ is of rank $N$. Thus, it is reasonable to assume that a set of $N-1$ independent fluctuation measurements can fully characterize entanglement in such a bipartitioned state. For example, take the case of two entangled qubits in an overall pure state. The entanglement spectrum is $\{p,1-p\}$ so that entanglement measures \cite{horodecki2001a,nielsen2001a,plenio2007a}, such as the von Neumann entanglement entropy $S = -\tr \rho_q \ln \rho_q$, purity $\gamma = \tr \rho_q^2$, and concurrence $C = \sqrt{4p(1-p)}$ \cite{wootters1998a}, depend on a single real parameter $0\le p\le 1$. For qubits, it is well-established that these entanglement measures are related to a subsystem fluctuation or variance. For instance, if we define an entanglement Hamiltonian as $H_q \equiv - \ln \rho_q$, a direct calculation leads to a fluctuation-entanglement relation of the form $\Delta^2 H_q = \mean{H_q^2} - \mean{H_q}^2 = C \ln \bigl(1+\sqrt{1-C^2}\bigr) - C\ln C$ \cite{feldman2009a,furman2015a}. 

In this paper we expand the notion of fluctuation-entanglement relations to the case of an arbitrary pure state partitioned into a $D$-state qudit and bath in such a way that the qudit is described by a reduced density operator of rank $N\le D$. Our main result is the identification of a set of $N-1$ operators that have fluctuations that collectively measure entanglement through a fluctuation-entanglement relation. We begin with a brief introduction to some notation and terminology, especially on the Bloch representation of reduced density operators $\rho_q$ as a linear combination of $\mathfrak{su}$($N$) generators $\{\Lambda_k\}$ and the identity operator (Section~\ref{generators}). We then express the variance of an $N-1$ component subsystem observable $\vec{w} \equiv \sum_k w_k \vechat{e}_k$ $(\vechat{e}_j\bm\cdot \vechat{e}_k = \delta_{jk})$ in terms of known entanglement measures like the linear entropy $S_L = 1-\gamma$ \cite{bartkiewicz2013a,buscemi2007a,peters2004a} and a generalized concurrence $C$ \cite{chen2005a} (Section~\ref{schmidtpolfluc}):
\begin{equation}
	\Delta^2 \vec{w} =2S_L = 2(1 - \gamma) = C^2.\label{flucent1}
\end{equation}
Here we refer to the total fluctuation $\Delta^2 \vec{w}$ as the Schmidt polarization fluctuation. This exact fluctuation-entanglement relation is true for any qudit-bath pure state and provides a particularly clear and simple mathematical basis for the intuitive link between entanglement and subsystem fluctuations, especially in a many-body context. Furthermore, this exact relation allows (i) recent experimental measurements of purity in many-body systems \cite{islam2015a} to be interpreted as direct measures of subsystem fluctuations and (ii) noise measurements (as in spin noise experiments \cite{ostereich2005a}) to be interpreted as direct entanglement measurements. We provide examples for cases when the subsystem partition is further decomposable into a product state (Section~\ref{sect:comp}) and when it is expressible in matrix product form (Section~\ref{sect:aklt}).

\section{Generators for $\mathfrak{su}(N)$}\label{generators}
A set of $N^2-1$ $\mathfrak{su}(N)$ generators that generalizes the $\mathfrak{su}(2)$ Pauli operators contains the following elements:
\begin{align}
{u}_{jk} & = \ket{j}\negthinspace\bra{k} + \ket{k}\negthinspace\bra{j},\\
{v}_{jk} & = -i\bigl( \ket{j}\negthinspace\bra{k} - \ket{k}\negthinspace\bra{j}\bigr),\\
{w}_{l} & = \sqrt{\frac{2}{l(l+1)}}\biggl[-l\ket{l+1}\negthinspace\bra{l+1} +\sum_{j=1}^l \ket{j}\negthinspace\bra{j}\biggr],
\end{align}
for $1\le j< k\le N$ and $1\le l\le N-1$ and $\{\ket{j}\}$ is an orthonormal set of $N$ state vectors that span a partition subspace \cite{hioe1981}. For notational convenience, these generators are often labeled as ${\Lambda}_1 = {u}_{12}, {\Lambda}_2 = {v}_{12}, {\Lambda}_3 = {w}_{1}, {\Lambda}_4 = {u}_{13},\dotsc, {\Lambda}_{N^2-1} = {w}_{N-1}$. These $N^2-1$ operators are traceless and satisfy $\tr{\Lambda_i\Lambda_j} = 2\delta_{ij}$. Furthermore, they are Hermitian and therefore represent observables on the corresponding partition. 

Any rank-$N$ reduced density operator ${\rho_q}$ can be written as a linear combination of the $\mathfrak{su}(N)$ generators $\Lambda_i$ and the $N$-dimensional identity $\mathbb{I}$:
\begin{equation}\label{bloch}
\rho_q = \frac{1}{N}\mathbb{I} + \frac{1}{2}\sum_{i=1}^{N^2-1} a_i \Lambda_i.
\end{equation}
The constants $a_i$ form the components of an $N^2 -1$ dimensional Bloch vector (or coherence vector) $\vec{a}$ \cite{hioe1981,jakobczyk2001,kimura2003a,byrd2003a}. The components of this Bloch vector are
\begin{equation}
a_j = \tr \Lambda_j\rho,
\end{equation}
which are expectation values of observables and thus experimentally accessible. Fluctuations in these observables have been used to construct a fluctuation-entanglement relation under the framework of generalized entanglement relative to the distinguished subset of observables $\{\Lambda_k\}$ \cite{barnum2003a,barnum2004a,somma2004a}. In this framework, one defines a vector $\bm{\Lambda} \equiv \sum_k \Lambda_k \vechat{e}_k$ so that the total component-wise fluctuation in $\bm\Lambda$ is
\begin{equation}
	\Delta^2 \bm\Lambda = \sum_{j=1}^{N^2-1} \tr \Lambda_j^2 \rho_q - \biggl(\sum_{j=1}^{N^2-1}\tr \Lambda_j\rho_q\biggr)^2.
\end{equation}
Each term in this expression is readily evaluated. The first term is the expectation value of the quadratic Casimir invariant $\sum_{j} \Lambda_j^2 = 2(N^2-1)\mathbb{I}/N$. The second term is the square of the Bloch vector, which can be expressed in terms of the purity $\gamma$ according to $\fabs{\vec{a}}^2 = 2(N\gamma-1)/N$ \cite{byrd2003a,kimura2003a}. We thus obtain a fluctuation-entanglement relation
\begin{equation}
  \Delta^2 \bm{\Lambda} = 2(N-\gamma).
\end{equation}
Calculating the fluctuation term in the left-hand side of this relation requires $N^2-1$ measurements. It contains the same information as the $N^2-1$ independent real parameters $a_j$ needed to completely specify the reduced density operator $\rho_q$ \eqref{bloch}. The main goal now in the remainder of this paper is to identify a set of $N-1$ fluctuation measurements that is equivalent to the set of $N-1$ independent eigenvalues of $\rho_q$. That is, we seek a set of fluctuations that corresponds to complete knowledge of the independent elements of the entanglement spectrum $\spec (\rho_q)$ \cite{li2008a}, thereby quantifying entanglement in an arbitrary pure state. Achieving this goal requires a set of $N-1$ compatible observables that commute with the reduced density operator $\rho_q$. A particularly convenient set is formed by the $N-1$ diagonal generators $w_k$ that are constructed from the Schmidt basis. In the following analysis, we find that the fluctuations in the non-diagonal generators $u_{jk}$ and $v_{jk}$ sum trivially in this basis.

\section{Schmidt polarization fluctuation}\label{schmidtpolfluc}

Let us consider a state vector $\ket{\psi}$ that is partitioned into subsystems $q$ and $B$ with Schmidt rank $N$:
\begin{equation}
\ket{\psi} = \sum_{j=1}^N \sqrt{p_j} \ket{j_q}\negthinspace\ket{j_B}.
\end{equation}
The set of state vectors $\{\ket{j_U}\}$ form an orthonormal basis  in the Hilbert subspace of partition $U$. Hence, the reduced density operator for subsystem $q$ is diagonal in this Schmidt basis:
\begin{equation}
\rho_q \equiv \tr_B \ket{\psi}\negthinspace\bra{\psi} = \sum_{j=1}^N p_j \ket{j_q}\negthinspace\bra{j_q}.
\end{equation}
Here the partial trace $\tr_B(\bm\cdot)$ is performed over the $\{\ket{j_B}\}$ states. For brevity, we will often omit the partition label subscripts on the state vectors as the reduced density operator is understood to act on the Hilbert subspace of a given partition. 

In the following discussion we define the Schmidt polarization fluctuation and obtain fluctuation-entanglement relations in terms of known entanglement measures. We begin with the simplest case of a qubit entangled with a bath before proceeding with the general qudit case. 

\subsection{Qubits}

In Schmidt decomposed form, the full density operator of a qubit $q$ entangled with an arbitrary environment $B$ is
\begin{equation}
	\rho  = p\ket{1_q}\negthinspace\ket{1_B}\negthinspace\bra{1_q}\negthinspace\bra{1_B} + (1-p) \ket{2_q}\negthinspace\ket{2_B}\negthinspace\bra{2_q}\negthinspace\bra{2_B},
\end{equation}
where $0\le p \le 1$. The reduced density operator for the qubit in the Schmidt basis $\{\ket{1},\ket{2}\}$ is therefore
\begin{equation}
	\rho_q = p \ket{1}\negthinspace\bra{1} + (1-p)\ket{2}\negthinspace\bra{2}.
\end{equation} 

It turns out that we can construct an entanglement measure that depends only on the fluctuation in the polarization of the qubit with respect to the Schmidt states $\ket{1}$ and $\ket{2}$. Let this polarization be represented by the diagonal $\mathfrak{su}(2)$ generator $w_1 = \ket{1}\negthinspace\bra{1} - \ket{2}\negthinspace\bra{2}$, which corresponds to the Pauli $z$-matrix in the Schmidt basis. We therefore have
\begin{equation}
	\mean{w_1} \equiv \tr w_1 \rho_q = 2p-1,
\end{equation}
and $\mean{w_1}$ measures the relative probability of finding the qubit in either of its Schmidt states: It is equal to $+1$ if the qubit is certainly in state $\ket{1}$, $-1$ if certainly in state $\ket{2}$, and zero when it is in either state with equal probability. Let us define the corresponding Schmidt polarization fluctuation as
\begin{equation}\label{Wqubit}
\Delta^{2} w_{1} = \tr w_1^2\rho_q - (\tr w_1\rho_q)^2 = 4p(1-p).
\end{equation}
It is evident that this fluctuation is a proper entanglement measure because it is a monotonic function of the qubit concurrence $C$ and, upon rearrangement, the linear entropy $S_L = 1-\gamma$:
\begin{equation}
	\Delta^{2} w_{1}= C^2 = 2[1-p^2 - (1-p)^2]= 2(1-\gamma) =2S_L .
\end{equation}
This simple result now sets the stage for a more general fluctuation-entanglement relation that applies to an $N$-level subsystem.

\subsection{Qudits}

The main challenge addressed in this paper is to generalize the previously derived relationship between the Schmidt polarization fluctuation and entanglement measures to the case of an entangled $D$-state qudit. To this end, we write the qudit ($q$) reduced density operator in terms of the Schmidt basis:
\begin{equation}
	\rho_q = \sum_{j=1}^{N} p_j \ket{j}\negthinspace\bra{j},
\end{equation}
where $N\le D$. That is, the dimension of the Hilbert subspace spanned by the Schmidt vectors is at most equal to the dimension of the full Hilbert space of the partition. We now define $N-1$ polarization observables in terms of the diagonal $\mathfrak{su}(N)$ generators so that 
\begin{equation}
	\mean{w_k} \equiv \tr{w_k \rho_q}.
\end{equation}
The generalized polarization expectation values are
\begin{align}
	\mean{w_1} &= p_1- p_2,\\
	\mean{w_2} &= \sqrt{\frac{1}{{3}}} \, ( p_1+ p_2 -2 p_3),\\
	\ &\ \,  \vdots \nonumber \\
	\mean{w_{N-1}} &= \sqrt{\frac{2}{N(N-1)}}\, \bigl[p_1 + p_2 + \dotsb - (N-1)p_N\bigr] .
\end{align}
Physically, the quantity $\mean{w_k}$ may be interpreted as a weighted relative probability of finding the qudit in one of the $j\le k$ Schmidt states $\{\ket{j}\}$ against finding it in the state $\ket{k}$. Furthermore, the polarization operators may be used to define a polarization vector  $\vec{w} \equiv \sum_k w_k \vechat{e}_k$, where $\{\vechat{e}_k\}$ is a set of $N-1$ orthonormal vectors. This construction allows the sum of polarization fluctuations to be written as
\begin{equation}
	\Delta^2 \vec{w} = \sum_{k=1}^{N-1} \Delta^2 w_k.
\end{equation}

We propose that the generalization of the qubit Schmidt polarization fluctuation $\Delta^2w_1$ in the case of a rank $N$ subsystem is the variance in the polarization vector:
\begin{equation}
\Delta^2 \vec{w}= \sum_{k=1}^{N-1} \bigl[\tr w_k^2\rho_q - (\tr w_k\rho_q)^2\bigr]. \label{Wqudit}
\end{equation}
This sum involves fluctuations in the $N-1$ diagonal generators $w_k$, which may be measured through appropriately generalized Stern--Gerlach-type experiments. To prove that the Schmidt polarization fluctuation $\Delta^2 \vec{w}$ is indeed a proper entanglement measure for the qudit-bath system, we demonstrate that it is a monotonic function of known entanglement measures. First, the sum of the mean squares of the polarizations is
\begin{equation}
	\sum_{k=1}^{N-1}\tr w_k^2\rho_q = \frac{2(N-1)}{N}.
\end{equation}
The sum of squares of the average polarizations takes some more effort to evaluate:
\begin{multline}
	\sum_{k=1}^{N-1}(\tr w_k\rho_q)^2 = \sum_{k=1}^{N-1}{\frac{2}{k(k+1)}}\biggl[k^2p_{k+1}^2 \\
	\quad - 2k p_{k+1}\sum_{j=1}^{k} p_j	 + \sum_{i=1}^k\sum_{j=1}^k p_ip_j\biggr].
\end{multline}
However, changing the summation order leads to simpler sums and we get
\begin{equation}
	\sum_{k=1}^{N-1}(\tr w_k\rho_q)^2 = \frac{2(N-1)}{N}\tr \rho_q^2 -\frac{4}{N} \sum_{1\le j < k \le N} p_jp_k.
\end{equation}
The latter term is simplified by the normalization condition $\sum_j p_j = 1$, which yields \cite{byrd2003a} 
\begin{equation}
	\sum_{1\le j < k \le N} p_jp_k = \frac{1-\tr \rho_q^2}{2}.
\end{equation}
Combining these results finally gives the desired fluctuation-entanglement relation between the Schmidt polarization fluctuation and known entanglement measures:
\begin{equation}\label{polflucent}
	\Delta^2 \vec{w} =2S_L = 2(1 - \gamma) = C^2.
\end{equation}
Here, $C$ is a generalized concurrence that has been reported in Ref.~\cite{chen2005a}. When the pure state is maximally entangled, the total Schmidt polarization fluctuation is equal to $2(N-1)/N$.

\section{Example: Decomposition over independent qudits}\label{sect:comp}

In the many-body context it may occur that the reduced density operator can be easily simplified into a product:
\begin{equation}
	\rho_q = \rho_q^{(1)} \otimes \rho_q^{(2)}.
\end{equation}
It is therefore important to determine how the Schmidt polarization fluctuations of independent subsystems relate to the total entanglement in composite subsystems. Given that the Schmidt polarization fluctuation of both components are known, 
\begin{equation}
	\Delta^2\vec{w}^{(j)} = 2(1-\gamma^{(j)}),
\end{equation}
we are able to construct the fluctuation-entanglement relation for the composite system. The total purity of $\rho_q$ is $\gamma = \gamma^{(1)}\gamma^{(2)}$ because purity is multiplicative over independent subsystems. The relationship between the total purity of the subsystem and the component Schmidt polarization fluctuations is therefore
\begin{equation}
	 \sum_{j=1}^2 \Delta^2\vec{w}^{(j)} - \tfrac{1}{2}\Delta^2\vec{w}^{(1)}\Delta^2\vec{w}^{(2)} = 2(1-\gamma).
\end{equation}
Since $\Delta^2\vec{w}^{(j)}/2 \le 1$, the polarization fluctuations are only approximately additive when higher order products of fluctuations are negligible. 

\begin{figure}[t!]
\centering
{\includegraphics[width=\linewidth]{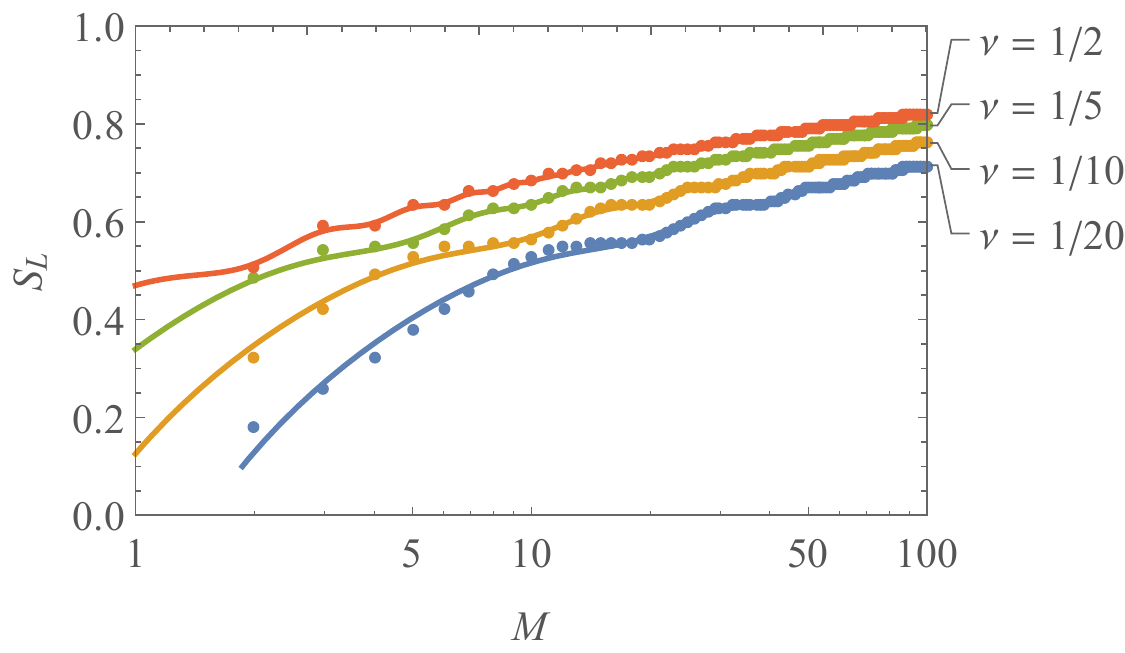}}
\caption{(Color online) The linear entropy $S_L$ of a block of $M$ adjacent sites in an infinite periodic chain occupied by non-interacting spinless fermions saturates logarithmically with $M$ to unity. The average fermion density per site is $\nu$. Data points are calculated from the fluctuation-entanglement relation \eqref{flucent:qudits} and the smooth curves are obtained from Fisher--Hartwig formulas with $O(M^0)$ corrections \cite{jin2004a,eisler2006a,abanov2011a,calabrese2011b}. } 
\label{fig:freefermions}
\end{figure}

In a decomposition with $M$ qudits, $\rho_q = \bigotimes_{j=1}^M\rho_q^{(j)}$, the fluctuation-entanglement relation becomes
\begin{equation}\label{flucent:qudits}
	\gamma = \prod_{j=1}^M\bigl[1-\tfrac{1}{2}\Delta^2\vec{w}^{(j)} \bigr].
\end{equation}
An important situation where such a decomposition is analytically tractable is encountered in the case of non-interacting spinless fermions hopping on a one-dimensional lattice \cite{chung2001a,cheong2004a,peschel2004a,peschel2009a}. In the pure ground state of this system, the reduced density operator corresponding to a block of $M$ adjacent sites can be written as a product of $M$ qubit density matrices. Thus, the identity \eqref{flucent:qudits} generalizes the determinant formulas for the entanglement entropy of free fermions \cite{jin2004a,calabrese2011a,calabrese2011b,calabrese2012a} that have been evaluated by the Fisher--Hartwig relation \cite{deift2011a}. As an illustration, we give the results of calculations of the block linear entropy $S_L$ in such a chain of fermions for several fermion densities $\nu < 1$ in Figure~\ref{fig:freefermions}. In this graph the component Schmidt polarization fluctuations are computed from
\begin{equation}
	\Delta^2\vec{w}^{(j)} = 4 p^{(j)}\bigl[1-p^{(j)}\bigr],
\end{equation}
where $p^{(j)}$ are the eigenvalues of the $M \times M$ matrix of fermion correlation functions $\mean{c_m^\dag c_n}$ with indices $m,n$ restricted to the block. As previously reported, entanglement is greatest at half-filling $\nu = 1/2$ and particle-hole symmetry leads to the linear entropy being equal for $\nu$ and $1-\nu$. Future work on the $M$ qudit formula \eqref{flucent:qudits} can extend this result to the case of multi-component atomic systems.

\section{Example: Ground state of the AKLT model}\label{sect:aklt}
Studies of entanglement in quantum spin chains have proved useful for the description of wires that can transmit quantum information between registers \cite{bose2007a} as well as for providing tractable theoretical models for investigating the relationship between entanglement and quantum critical phenomena \cite{vidal2003a,cui2012a,franchini2014a,fan2014a}. 
As an example, we consider the valence-bond solid (VBS) ground state of the spin-1 Affleck-Kennedy-Lieb-Tasaki (AKLT) model \cite{aklt1987}:
\begin{equation}
	H = \sum_{j=1}^L \vec{S}_j\bm\cdot \vec{S}_{j+1} + \frac{1}{3} (\vec{S}_j\bm\cdot \vec{S}_{j+1} )^2. 
\end{equation} 
The VBS ground state is separated from excited states by a Haldane gap \cite{affleck1989a} and the entanglement entropy saturates for long blocks. We impose periodic boundary conditions on the spin-1 operators $\vec{S}_{L+1} = \vec{S}_1$ and take the thermodynamic limit $L\to\infty$. The reduced density operator $\rho_\ell$ for a contiguous block of $\ell$ spins can be calculated from the matrix product representation of the VBS state \cite{santos2012a}. Although the full Hilbert space for the block is $3^\ell$-dimensional, the reduced density operator is at most rank-4. Its eigenvalues are
\begin{align}
	p_1 &= p_2 = p_3 = \frac{1-(-3)^{-\ell}}{4},\\
	p_4 &= \frac{1+3(-3)^{-\ell}}{4}.
\end{align}
The fluctuation in the Schmidt polarization vector is
\begin{equation}
	\Delta^2 \vec{w} = \sum_{k=1}^3 \Delta^2 w_k = \tfrac{3}{2} (1-9^{-\ell}),
\end{equation}
where the relevant polarization operators in the Schmidt basis are
\begin{gather}
	w_1 = \begin{pmatrix}
		1&	& &\\
		& -1& &\\
		& & 0&\\
		& & & 0
	\end{pmatrix}, \quad w_2 = \frac{1}{\sqrt{3}}\begin{pmatrix}
		1&	& &\\
		& 1& &\\
		& & -2&\\
		& & & 0
	\end{pmatrix}, \nonumber \\
	w_3 = \frac{1}{\sqrt{6}}\begin{pmatrix}
		1&	& &\\
		& 1& &\\
		& & 1&\\
		& & & -3
	\end{pmatrix}.
\end{gather}
The fluctuation-entanglement relation \eqref{polflucent} therefore gives the subsystem purity as
\begin{equation}
 \gamma= \tfrac{1}{4}{(1 +  3^{1-2\ell})} ,
\end{equation}
which converges exponentially fast with $\ell$ to the maximally entangled value $\gamma_\text{max} =1/4$ \cite{fan2004a}.

\section{Concluding remarks}
Subsystem fluctuations have been studied as probes of  entanglement, with notable results for many-body systems \cite{klich2009a,song2010a,song2012a,calabrese2012a}. Still, only few exact and general relationships have been derived between an entanglement entropy and a subsystem variance in the many-body situation. Specific results have been reported for the case of independent qubits \cite{klich2006a,purkayastha2014a,puspus2014a,gigena2015a}, one-dimensional free fermions \cite{calabrese2012a}, and Luttinger liquids \cite{song2010a,song2012a,petrescu2014a}. 

In this work we have derived a model-independent fluctuation-entanglement relation that provides a general, yet simple, quantitative basis to the long-held assertion that subsystem fluctuations can measure entanglement. Our result implies that it takes $N-1$ fluctuation measurements to construct a fluctuation-based entanglement measure \eqref{polflucent} for a rank-$N$ reduced density operator. In fact, since the fluctuation $\Delta^2w_k$ depends on the first $k$ eigenvalues of the reduced density matrix, knowledge of these $N-1$ fluctuations is equivalent to knowledge of the full entanglement spectrum $\{p_k\}$. It therefore becomes increasingly cumbersome to use such a measure for a general entangled many-body system with exponentially large Schmidt rank. However, there are broad classes of many-body states that have reduced density matrices with well-bounded rank as in the examples presented here: models mappable onto one-dimensional free fermions \cite{peschel2004a,peschel2009a}, and matrix product states and the states that are well-approximated by them \cite{fannes1992a,verstraete2006a}.

\section{Acknowledgments}
A. V. acknowledges support from the DOST Science Education Institute through its ASTHRD Program. The authors thank J.~P.~Esguerra, E.~A.~Galapon, and K.~H.~Villegas for helpful discussions.


\end{document}